# Hole-Induced Electronic and Optical Transitions in La$_{1-x}$Sr$_x$FeO$_3$ Epitaxial Thin Films

Le Wang[1], Yingge Du[1*], Peter V. Sushko[1], Mark E. Bowden[2], Kelsey A. Stoerzinger[1], Steven M. Heald[3], Mark. D. Scafetta[1], Tiffany C. Kaspar[1], Scott. A Chambers[1*]

[1]Physical and Computational Sciences Directorate, Pacific Northwest National Laboratory, Richland, Washington 99354, USA

[2]Environmental Molecular Sciences Laboratory, Pacific Northwest National Laboratory, Richland, Washington 99352, USA

[3]Advanced Photon Source, Argonne National Laboratory, Lemont, Illinois 60439, USA

[*]Authors to whom correspondence should be addressed: yingge.du@pnnl.gov and scott.chambers@pnnl.gov.

We have investigated the electronic and optical properties of epitaxial La$_{1-x}$Sr$_x$FeO$_3$ for $0 \leq x \leq 1$ prepared by molecular beam epitaxy. Core-level and valence-band x-ray photoemission features monotonically shift to lower binding energy with increasing x, indicating downward movement of the Fermi level toward to the valence band maximum. Both Fe 2$p$ and O 1$s$ spectra broaden to higher binding energy with increasing x, consistent with delocalization of Sr-induced holes in the Fe 3$d$ – O 2$p$ hybridized valence band. Combining X-ray valence band photoemission and O $K$-edge x-ray absorption data, we map the evolution of the occupied and unoccupied bands and observe a narrowing of the gap, along with a transfer of state density from just below to just above the Fermi level, resulting from hole doping. In-plane transport measurements confirm that the material becomes a $p$-type semiconductor at lower doping levels and exhibits a insulator-to-metal transition at x = 1. Sub-gap optical transitions revealed by spectroscopic ellipsometry are explained based on insight from theoretical densities of states and first-principles calculations of optical absorption spectra.

## I. INTRODUCTION

Perovskite (ABO$_3$-type) transition-metal oxides have been of significant interest over the past thirty years because of their intriguing properties. These include metal-insulator transitions (MIT) [1], colossal magnetoresistance (CMR) [2], high-temperature superconductivity [3], and ferroelectricity [4]. These materials have also been of interest for applications in renewable energy technologies including electrochemical and photoelectrochemical water splitting [5-7], photovoltaic solar cells [8,9], and solid oxide fuel cells [10,11]. LaFeO$_3$ (LFO), is a wide-gap antiferromagnetic insulator (E$_g$~2.3 eV), [12] whereas stoichiometric SrFeO$_3$ (SFO) is a helical antiferromagnetic metal [13]. As a result, the La$_{1-x}$Sr$_x$FeO$_3$ (LSFO) solid solutions are of interest for solar water splitting because the (in principle) tunable band gap can enhance light absorption across the visible portion of the solar spectrum [14-16]. It has been shown that LFO is orthorhombic at room temperature. However, increasing the Sr concentration alters the crystal structure to rhombohedral near x = 0.3 and to cubic near x = 0.8 [17,18]. In addition, hole doping by Sr$^{2+}$ substitution for La$^{3+}$ introduces *p*-type conductivity and enhances the electrocatalytic oxygen evolution reaction (OER) [19,20]. In our previous work, we found that Sr doping lowers the optical bandgap and increases the photovoltage under 460 nm blue light emitting diode (LED) illumination by a negligibly small value for pure LFO, and to more than 0.5 V for La$_{0.8}$Sr$_{0.2}$FeO$_{3-\delta}$ [21]. Determining how the electronic and optical properties evolve in LSFO is thus critical to understanding and controlling the OER activity in this system.

Sr$^{2+}$ substitution for La$^{3+}$ should dope holes into the Fe-O hybridized valence band [22,23], resulting in an increase in the nominal Fe valence from 3+ to 4+ [24]. However, LSFO films are often found to be oxygen deficient as the formation energy of oxygen vacancies (V$_O$) is low [25,26], leading to hole compensation, and the Fe valence remains closer to 3+. As shown by Xie *et al.* [26], the electronic structure and transport properties of LSFO films are very sensitive to the oxygen stoichiometry. Post annealing of films grown by molecular beam epitaxy (MBE) is critical to revealing the instrinsic properties of LSFO [27]. Wadati *et al.* [28] investigated the electronic structure of LSFO films by X-ray photoemission spectroscopy (XPS) and X-ray absorption spectroscopy (XAS). These authors found that the valence band (VB) spectral feature nearest to Fermi level (E$_F$) becomes weaker and moves toward E$_F$ as x was increased. However, the maximum x value they explored was 0.67. She *et al.* [20] reported Fe 2*p* spectra for LSFO (0≤x≤1) power samples, but the binding enengy scale was not calibrated and the valence bands were not directly measured. Scafetta *et al*. [24,29] published a careful investigation on the optical properties of LSFO which revealed that increasing x induces systematic changes in the optical spectra, including a red-shift of the absorption features at energies below the LFO band gap. However, due to the limited energy range of the ellipsometer used in those studies (from 1.2 to 5 eV), sub-eV optical transitions could not be measured.

In this study, we seek to fill in the gaps in existing knowledge of this system by investigated a set of well-defined, epitaxial $La_{1-x}Sr_xFeO_3$ films over the full range of x, combining x-ray photoemission and absorption spectroscopies with theory, and using spectroscopic ellipsometry from the infra-red to the ultraviolet to accurately map and explain densities of states. All films were grown on (001)-oriented $SrTiO_3$ (STO) substrates by oxygen plasma assisted molecular beam epitaxy (OPA-MBE). Here activated oxygen was used to maximize the extent of LSFO film oxidation. By combining *in situ* core-level Fe 2*p* and valence band photoemission spectra, *ex situ* Fe and O *K*-edge x-ray absorption, in-plane transport and optical absorption, we develop a more complete understanding of the evolution of the functional properties of LSFO. We also reveal the origins of the low-energy optical absorption bands by carrying out time-dependent density functional theory (TDDFT) simulations.

## II. METHODS

Epitaxial $La_{1-x}Sr_xFeO_3$ films with x = 0, 0.2, 0.37, 0.5, 0.8, and 1.0 and thicknesses of 10-30 nm were grown on undoped $TiO_2$-terminated $SrTiO_3$(001) (STO) substrates by OPA-MBE. La, Sr, and Fe were evaporated from high temperature effusion cells. Evaporation rates were calibrated prior to film growth using a quartz crystal oscillator. The substrate temperature was 700°C, and the oxygen partial pressure was $\sim 2\times10^{-6}$ Torr. *In situ* reflection high-energy electron diffraction (RHEED) was used to monitor the overall growth rate and surface structure. After film growth, the sample temperature was lowered to ambient at a rate of 30 °C/min in $\sim 2\times10^{-6}$ Torr of activated oxygen from an electron cyclotron plasma source. In order to obtain fully oxidized LSFO, the as-grown films were post-annealed by heating to 600°C in activated oxygen at a chamber pressure of $\sim 3\times10^{-5}$ Torr for one hour. They were then cooled to 200°C at a rate of 5°C/min$^{-1}$, held at this temperature for two hours, then further cooled at 1°C/min until the heater reached zero output power. Finally, the films were kept in the activated oxygen stream at ~3x10$^{-5}$ Torr for another hour [30].

High-resolution XPS excited with monochromatic Al-K$\alpha$ x-rays were measured at normal emission with a VG/Scienta R3000 electron energy analyzer in an appended chamber. The energy resolution was 0.50 eV. Spectra were measured immediately after film growth using an electron flood gun to compensate the positive photoemission charge as these films were not all conductive, and those that were conductive were not grounded. Samples were then removed from UHV and a polycrystalline Au foil was placed in direct electrical contact with the film surface and grounded to the sample holder. The Au $4f_{7/2}$ peak was used to calibrate the binding energy scale. XPS measurements were then repeated after oxygen plasma cleaning in the MBE chamber. Some residual contamination remained on the film surfaces as evidenced by weak features on the higher binding energy side of the O 1$s$ lattice peaks, along with a weak C 1$s$ peak. Near-Edge X-ray Absorption Fine Structure (NEXAFS) data were collected in electron yield mode at Beamline 9.3.2 at the ALS. A linear pre-edge background was subtracted from the spectra, followed by normalization to the post-edge and five-point smoothing. The latter did not alter the spectral features in any significant way. Fe K-edge XANES was measured at the APS.

Film crystallography was investigated by high-resolution X-ray diffraction (HRXRD, Philips X'Pert diffractometer) with Cu K$_{\alpha 1}$ radiation ($\lambda$ = 1.5406 Å). Spectroscopic ellipsometry measurements were performed using a rotating analyzer based instrument with a compensator (V-VASE; J.A. Woollam Co., Inc.) over the spectral range from 0.36 to 4.76 eV. Spectra were collected at three different incident angles (65°, 70° and 75°). A simple two-layer model was used to extract refractive indices ($n$) and extinction coefficients ($k$). The inter-band optical transitions were modelled as direct, dipole-forbidden excitations [12,19]. To this end, the linear regions in the quantities ($\alpha E$)$^{2/3}$ (where $\alpha$ is the absorption coefficient and $E$ is the photon energy) were extrapolated to the photon energy axis. In-plane transport properties were investigated using a Hall measurement system in the temperature range 100−310 K. Measurements were done in the van der Pauw geometry with square samples (~5×5 mm$^2$) and silver paste in the corners.

Optical absorption spectra were calculated from first principles using a two-step procedure. First, structural parameters were determined using the periodic model approach and density functional theory (Fig. S1). We used a 40 atom (2×2×2) orthorhombic perovskite cell La$_8$- $_{n}Sr_{n}Fe_{8}O_{24}$ where 8-$n$ La and $n$ Sr atoms were randomly distributed throughout the A sublattice. For simplicity, we considered one configuration for each Sr concentration. The geometrical structures were optimized using PBEsol functional [31], as implemented in the Vienna Ab initio Simulation Package (VASP) [32]. The projector augmented-wave was used to approximate the electron-ion potential [33]; the 4×4×4 Mokhorst-Pack $k$-grid was used for the Brillouin zone integration. The plane wave basis cutoff was set at 500 eV and the total energy was converged to $10^{-6}$ eV. Unless stated otherwise, all geometrical structures were fully relaxed.

In the second step, the pre-optimized LSFO solid solution lattices were modeled using an embedded cluster approach [34-36]; excitation energies and corresponding transition probabilities were calculated using TDDFT [37] as implemented in NWChem [38]. In the embedded cluster approach, a part of the lattice is treated quantum mechanically (referred to as the quantum cluster) and is embedded into the electrostatic and short-range potential generated by a surrounding lattice (referred to as the environment). To this end, we complemented the pre-optimized periodic model structures with auxiliary point charges, as described elsewhere [39,40], in order to eliminate all components of the low multipole moments from $m = 1$ (dipole) to $m = 4$ (hexadecapole) without affecting the charge neutrality condition. The modified LSFO supercells were used to generate spherical nanoclusters (one for each periodic structure) of radius ($R$) equal to ~5.7 nm. The electrostatic potential in the central region of such clusters converges absolutely with increasing $R$. Its variations are less than 0.01 V within the quantum cluster for $R$ = 5.7 nm (see Fig. S1 in the Supplementary Information section).

The quantum clusters were constructed so as to capture electronic structure changes associated with a growing Sr concentration. In all cases, these clusters include eight Fe ions and all neighboring oxygen ions as well as seven A-site ions, these being either $La^{3+}$ or $Sr^{2+}$ (See Fig. S1). Cations adjacent to the quantum cluster are represented using effective core pseudopotentials that mimic the radius and ionic charges of Fe, Sr, and La. The details of the basis sets are described in the Supplemental Information. The ground state electronic structures of these clusters were calculated using B3LYP functional [41] in the open-shell mode. Knowledge of the atomic orbital contributions to cluster orbitals and an ability to select occupied and virtual orbitals for the TDDFT active space allows us to assign contributions of various types of transitions to specific features in the experimental absorption spectra.

## III. EXPERIMENTAL CHARACTERIZATION

### A. Structure and Film Surface Quality

Figure 1(a) shows RHEED patterns for the as-grown LSFO films. These patterns exhibit sharp, unmodulated streaks, revealing excellent crystallinity and flat surfaces. Typical AFM images [Fig. 1(b)] for the as-grown LSFO films show clearly spaced surface steps, similar to those seen on the STO substrates. X-ray reciprocal space maps (Fig. S2) of the film and substrate (103) peaks reveal that the LSFO films are fully strained. The (002) peak of the LSFO film systematically shifts to higher angle with increasing x, as shown in Fig. 1(c), indicating that the out-of-plane lattice parameter decreases (Table 1 and Fig. S2). Well-defined Kiessig fringes are visible in Fig. 1(b), confirming the high degree of crystallinity in these fully-strained films.

### B. Core-level Photoemission Spectra

Figure 2(a) shows representative core-level spectra for the four elements in LSFO as a function of x. The Fe $2p$ and O $1s$ peaks become progressively more asymmetric on their high-binding energy sides with increasing x, indicating significant hybridization of O $2p$ and Fe $3d$ orbitals in the VB and hole delocalization therein. In contrast, the Sr $3d$ and La $4d$ line shapes do not change with x. The low-binding-energy side of the O $1s$ lattice peaks is unaffected by residual surface contamination resulting from removal from UHV for sample grounding (Fig. S3). We thus use this portion of the spectrum to monitor the relationship between core level binding energy shifts and hole doping. A common shift to lower binding energy is measured for the O $1s$, Sr $3d$, La $4d$ core levels due to the change in chemical potential with x [28,42,43]. We average these three energy shifts to track this change. Figure 2(b) shows the chemical potential trend and indicates a monotonic downward shift with hole concentration, similar to that seen in other hole-doped transition oxides alloys, such as $La_{1-x}Sr_xTiO_3$ [44], $La_{1-x}Sr_xMnO_3$ [45], and $La_{1-x}Sr_xCrO_3$ [42]. The trend in chemical potential nicely matches the trend in valence band maximum (VBM) shown in Fig. S3 and Table 1, as well as the decreases in both the room-temperature resistivity [Fig. 2(c) and Fig. S4] and hopping activation energy [27].

The change in Fe valence with Sr doping is clearly seen in Fe $2p$ XPS and Fe $K$-edge XAS spectra in Fig. 3. In order to more easily visualize the change in Fe $2p$ lineshape with x, we align all XPS so the corresponding O $1s$ peaks are at 530.0 eV in order to remove the chemical potential change with x. The results are shown in Fig. 3(a). For pure LFO, the Fe 2*p* line shape is virtually identical to that in binary oxides containing $Fe^{3+}$ such as γ-$Fe_2O_3$(001) [46]. However, the Fe $2p_{3/2}$ and Fe $2p_{1/2}$ peaks become broader and shift to higher binding energy with increasing x, suggestive of progressive partial hole localization a on B-site Fe cations. Moreover, the charge transfer satellites become weaker with increasing x and are not observed for SFO (x = 1). The spectrum of SFO is very similar to that in the fully oxidized $BaFeO_3$ thin films [47], again consistent with hole doping on Fe sites, resulting in a formal charge larger than 3+. (The extent of Fe valence increase depends on the amount of Fe 3*d* – O 2*p* hybridization in the VB.) This trend is corroborated by the Fe *K*-edge XAS data (Figure 3(b)). A systematic shift of the maximum absorption feature to higher x-ray energy with increasing x is clearly seen, again indicating an increase in formal charge. The increase in Fe valence should result in a decrease in the ionic radius, leading to an expected decrease in the (unconstrained) out-of-plane lattice parameter. This trend is clearly seen in the XRD results (see Fig. S2 and Table 1). Moreover, the small pre-edge feature in the Fe *K*-edge XAS also increases with x, indicating an increase in hole concentration in the Fe 3*d*-derived VB, consistent with previous results [48]. Significantly, we did not observe a leveling off of either the absorption peak maximum energy shift or the intensity of the pre-edge feature as x approaches 1, as noted in earlier work [48]. We attribute this to a higher oxygen content in our films, particularly SFO.

### C. Electronic and Optical Properties

To gain deeper insight into the relationship between hole doping and the optical properties of LSFO films, we correlate optical absorption (OA) spectra with electronic densities of states. Figure 4(a) shows OA spectra for the LSFO film series. Characteristic features in the low-energy portion of the spectra are denoted as A and B. We show in Fig. 4(b) occupied and unoccupied densities of states taken from XPS VB and O *K*-edge XAS, respectively. The energy scale for the XAS spectra was aligned so the band gaps agree with those from the OA measurements [Fig. 4(a)]. To interpret the origin of features A and B, we also calculate the one-electron DOS and the theoretical OA spectra for LSFO. Figures 4(d) show the DOS from an embedded cluster model with the B3LYP density functional. The distinct feature at the top of the VB in pure LFO [Fig. 4(b)] consists of hybridized O 2*p* and Fe 3*d* $e_g$ orbitals, while unoccupied Fe 3*d* $t_{2g}$ and $e_g$ orbitals form the bottom of conduction band (CB) [22,28]. Upon hole doping, the top of the VB becomes partially unoccupied and a new DOS feature appears above $E_F$ (marked by an arrow in the x =

0.2 XAS spectrum in Fig. 4(b), resulting in a narrowing of the gap. Previous reports also assign the new feature above $E_F$ to a split-off, empty Fe $3d$ $e_g$/O $2p$ hybridized band, consistent with our calculations [22,24,28]. Therefore, we attribute feature B in the OA [Fig. 4(a)] to transitions from the top of VB to the split-off empty Fe $3d$ $e_g$/O $2p$ hybridized band in the gap, as seen schematically in Fig. 4(c). The intensity of feature B increases and its energy drops with increasing x [Fig. 4(a)], indicating the growth of the unoccupied DOS just above $E_F$ and a closing of the gap due to hole doping.

In assigning feature A, we introduce the following notation based on LFO being a G-type antiferromagnet. Each B-site $Fe^{3+}$ ion with spin-up (↑) $3d$ occupied ($o$) orbitals $e_g$↑$o$ and $t_{2g}$↑$o$ as well as virtual ($v$) spin-down (↓) orbitals $e_g$↓$v$ and $t_{2g}$↓$v$, is adjacent to an $Fe^{3+}$ ion with the opposite spin orientation ($e_g$↓$o$ and $t_{2g}$↓$o$ and $e_g$↑$v$ and $t_{2g}$↑$v$). Using these notations, we assign feature A to the spin-allowed transitions between neighboring sites, O 2p/Fe $e_g$↑$o$ → $t_{2g}$↑$v$ and O 2p/Fe $e_g$↓$o$ → $t_{2g}$↓$v$, as shown schematically in Fig. 4(c). Their onset energies can be determined from Tauc plots, as seen in the inset of Fig. 4(a) and Table 1. Feature A becomes stronger and shifts to lower energy with increasing x. This shift reveals that the separation between the valence band and the band derived from the virtual $t_{2g}$ orbitals decreases with increasing x (Fig. S3).

These assignments are supported by first-principles calculations of the OA spectra. Theoretical spectra [Fig. 4(e)] were generated by calculating excitation energies and oscillator strengths for the 600 lowest-energy transitions that correspond to features A (green) and B (red). Each transition was represented using a Gaussian function with a full width at half maximum (FWHM) value of 0.1 eV. We note that oscillator strengths for optical transitions are sensitive to the details of the local environment, including lattice distortions due to alloy formation and vibrational amplitudes. In order to accurately account for these effects, we would need to average over all actual distributions of La and Sr atoms within the A sublattice, as well as over thermally induced atomic motion [15,35]. Here we mimic such averaging by assigning the same oscillator strength to each transition, with only one La/Sr distribution in the A sublattice considered for each x value. The accuracy of this approximation can be evaluated by comparison of the calculated and observed OA spectrum for LFO [see Fig. 4(e)]. The low-energy peak at ~4 eV and the absorption increase at ~4.5 eV are consistent with the experimentally observed features at 3.2 and 3.7 eV, respectively. Analysis of the molecular orbitals contributing to these transitions reveals that they are formed by hybridized Fe $e_g$/O $2p$ → $t_{2g}$-$v$ excitations, as expected from the structure of the DOS, while the higher energy peak has an additional contribution due to excitations from pure O $2p$ orbitals located deeper in the VB to the $t_{2g}$-$v$ orbitals. We note that the onset of these transitions shifts to lower energies, as seen in Fig. 4(e), in agreement with the experimental data [Fig. 4(a)].

As mentioned above, the transfer of DOS from the top of the VB (occupied) to just above $E_F$ (unoccupied) resulting from hole doping creates new low-energy optical transitions denoted as B in Fig. 4(a) and Fig. 4(c). The experimentally determined onset energy of this feature approaches zero with increasing x. This trend is consistent with the vanishing band gap in the periodic model simulations using PBEsol density functional [Fig. S1(c)]. However, reproducing a vanishing gand gap is problematic for the hybrid B3LYP functional due to the nature of its exchange-correlation term. Hence, our calculations do not reproduce the systematic shift of the onset of feature B with increasing x. However, the gradual increase in intensity of this feature is visible in both experimental and theoretical data. We note that transitions of the kind VB → Fe $e_g$/O $2p$ transitions also contribute to feature A. While our data do not allow us to isolate these two contributions to feature A, we propose this could be achieved via mapping feature A as a function of strain, temperature and oxygen content.

## V. CONCLUSIONS

We show that $Sr^{2+}$ substitution for $La^{3+}$ in epitaxial $La_{1-x}Sr_xFeO_3$ thin films deposited on STO(001) increases the electrical conductivity as a result of hole doping of the VB, leading to an insulator-to-metal transition at x = 1. The Fermi level drops toward the top of VB with increasing x. These changes in electronic structure are manifested by the formation of a new split-off band above the Fermi level seen in both experiment and first-principles calculations. The evolution of the low-energy optical absorption peaks can be explained by the changes in electronic structure in LSFO accompanying hole doping. Our calculations reveal that for LFO, the top of VB consists of hybridized O $2p$ and Fe $3d$ $e_g$ orbitals, whereas unoccupied Fe $3d$ $t_{2g}$ and $e_g$ orbitals form the bottom of CB. The optical absorption feature at ~2.5 – 3 eV is assigned to the spin-allowed transitions O $2p$/Fe $e_g\uparrow o$ → $t_{2g}\uparrow v$ and O $2p$/Fe $e_g\downarrow o$ → $t_{2g}\downarrow v$, where the $e_g$ and $t_{2g}$ orbitals are localized on adjacent Fe ions. The new feature that forms just above the Fermi level is assigned to the split-off empty Fe 3$d$ $e_g$/O 2$p$ hybridized band resulting from hole doping. This feature gives rise to a new set of optical transitions from the VB at ~1 eV. The enhanced level of understanding of the electronic properties and optical transitions in LSFO presented here can facilitate the usefulness of LSFO in photocatalysis, photovoltaics, and electrocatalysis.

## ACKNOWLEDGMENTS

This research was supported by the U.S. Department of Energy (DOE), Office of Basic Energy Sciences, Division of Materials Science and Engineering under award No. 10122. This research used resources of the Advanced Light Source, which is a DOE Office of Science User Facility under contract no. DE-AC02-05CH11231. O $K$-edge measurements and anlysis was supported by the Linus Pauling Distinguished Post-doctoral Fellowship at Pacific Northwest National Laboratory (PNNL LDRD 69319). A portion of the work was performed at the W. R. Wiley Environmental Molecular Sciences Laboratory, a DOE User Facility sponsored by the Office of Biological and Environmental Research. PNNL is a multi-program national laboratory operated for DOE by Battelle.

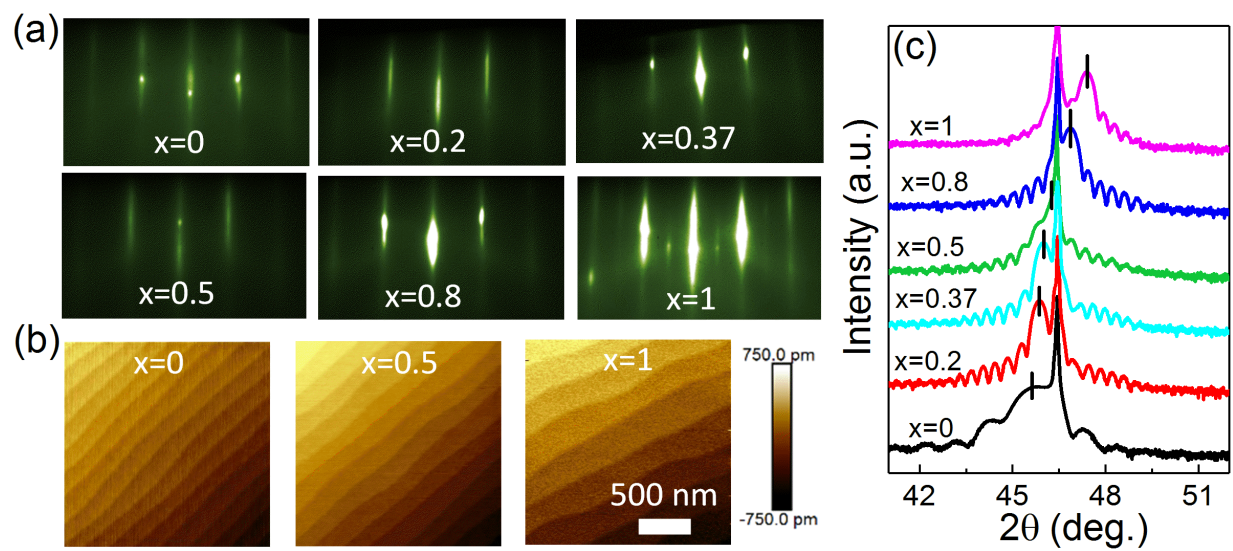

FIG. 1. (a) RHEED patterns for La$_{1-x}$Sr$_x$FeO$_3$ films grown on STO(001) viewed along the [100] zone axis. (b) Representitive AFM images for LSFO films with x = 0, 0.5 and 1 on STO(001). The scan image size is 2 μm × 2 μm. (c) XRD θ-2θ scans near the (002) peak for the LSFO/STO(001) film set. All LSFO films were annealed in activated oxygen to maximize the extent of oxidation.

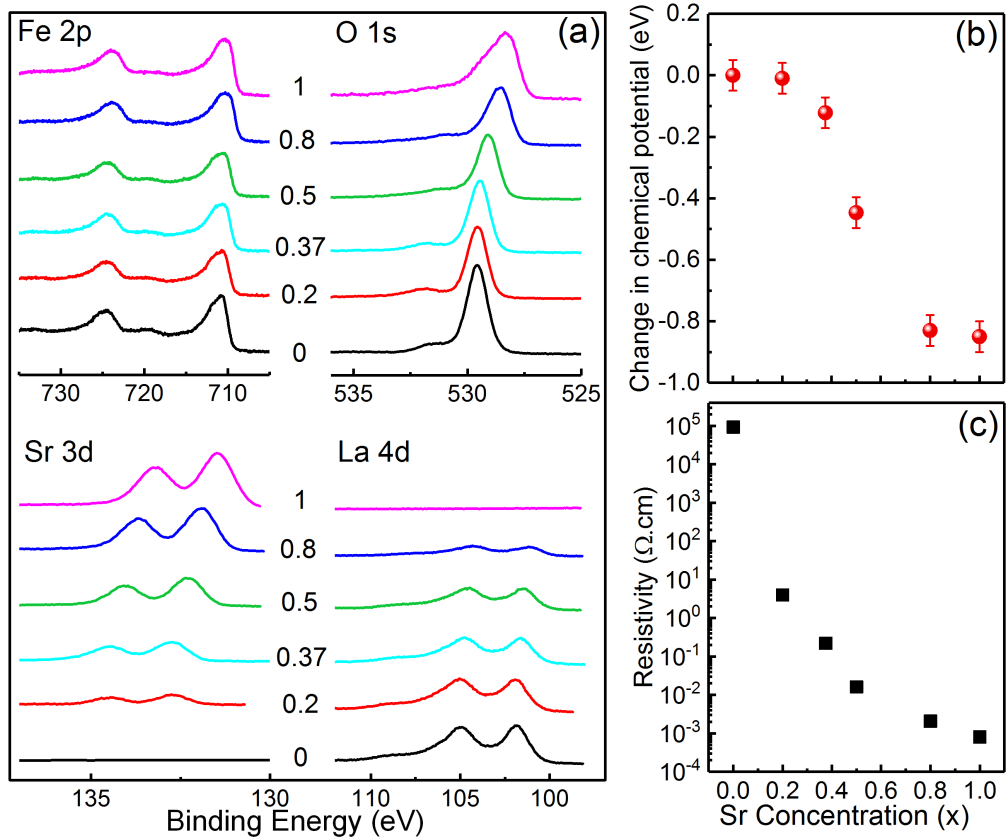

FIG. 2. (a) Fe $2p$, O $1s$, Sr $3d$ and La $4d$ XPS for La$_{1-x}$Sr$_x$FeO$_3$ as a function of x. All films were electrically grounded and the binding energy scales were calibrated using the Au $4f_{7/2}$ peak binding energy (84.00 eV). (b) Average chemical potential shift deduced from O $1s$, Sr $3d$ and La $4d$ binding energies. (c) Room temperature resistivity vs x.

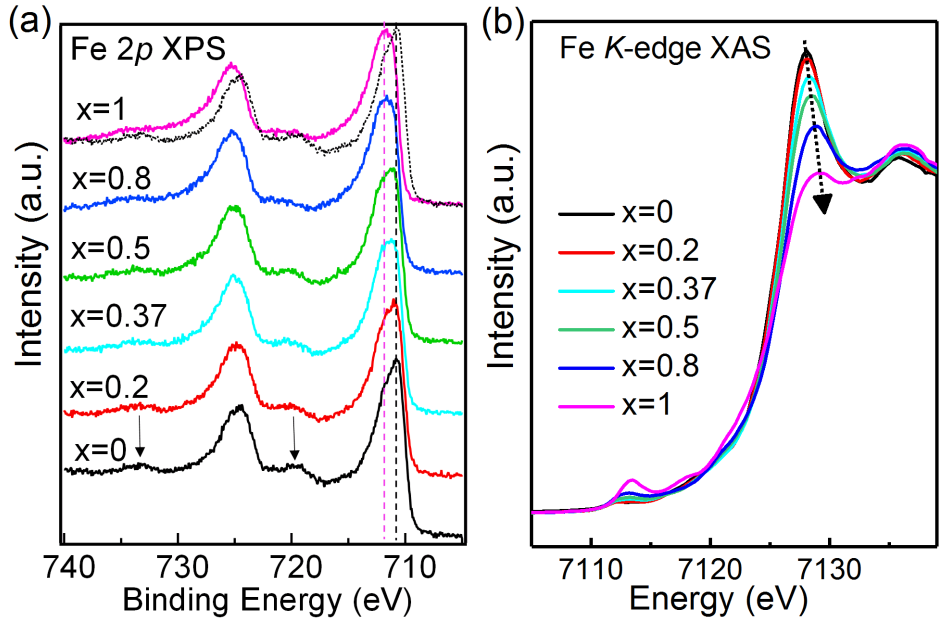

FIG. 3. (a) Fe $2p$ core level XPS for $La_{1-x}Sr_xFeO_3$ as a function of x. All spectra were shifted so the associated O $1s$ peaks fall at 530.0 eV. As a guide to the eye, we mark the core-level peaks of the LFO and SFO films with dashed lines. The satellite peaks for $Fe^{3+}$ (marked by the arrows) are absent in $SrFeO_3$ ($Fe^{4+}$). (b) Fe $K$-edge XAS for $La_{1-x}Sr_xFeO_3$ as a function of x. The dash arrow indicates the peak shift.

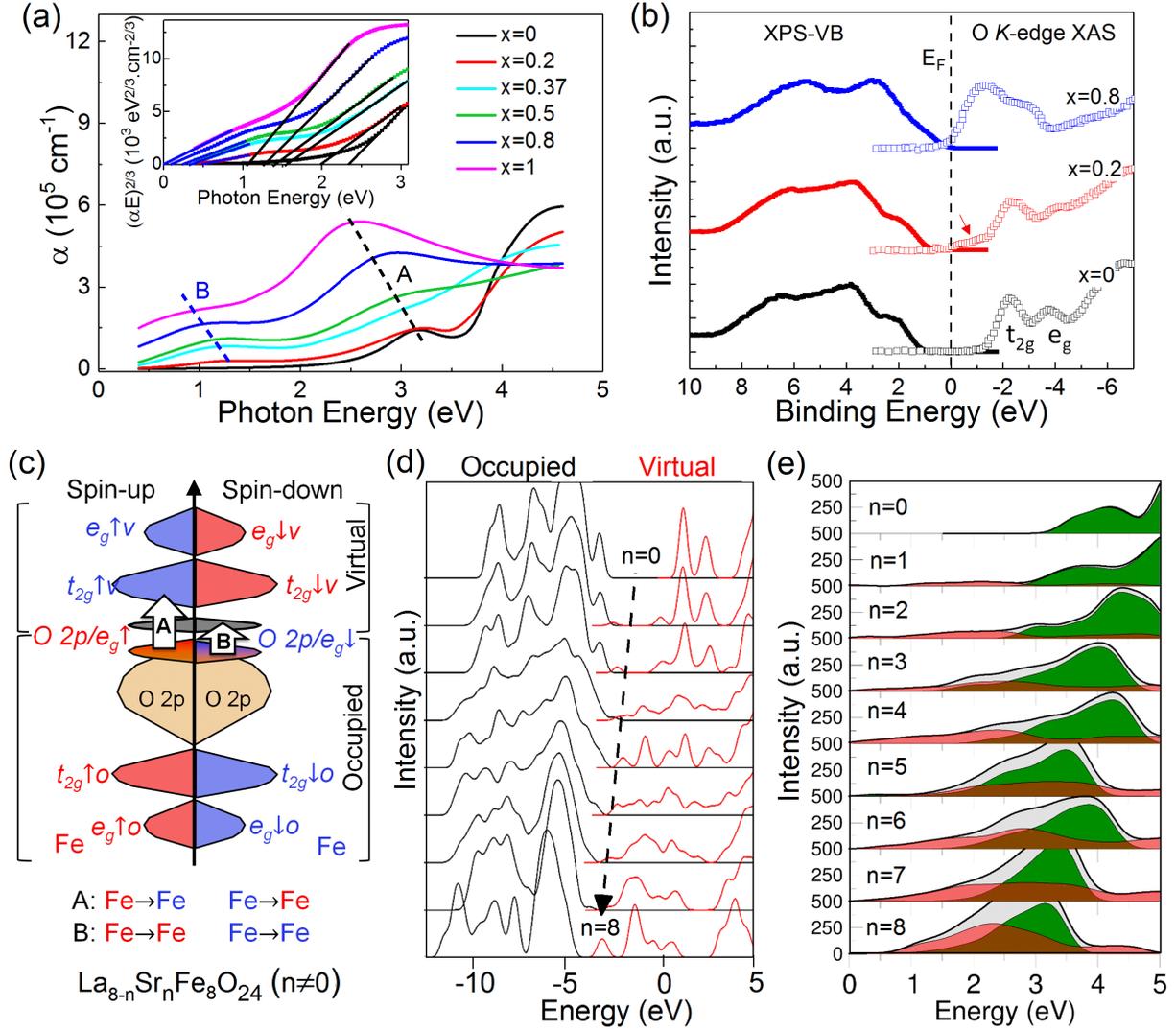

FIG. 4. (a) Optical absorption spectra for $La_{1-x}Sr_xFeO_3$ films deposited on STO(001). The inset shows plots of $(\alpha E)^{2/3}$ versus photon energy E for the direct dipole-forbidden excitations. The onsets of transitions A and B are summarized in Table 1. (b) XPS-VB (left) and O $K$-edge XAS (right) for selected LSFO films. (c) Schematic of the electronic structure and optical transitions for $La_{8-n}Sr_nFe_8O_{24}$ (n≠0, equivalent to x>0 in $La_{1-x}Sr_xFeO_3$). Blue and red colors refer to neighboring Fe atoms that carry up and down spin-density, respectively. Hybridization of O $2p$ and Fe $e_g$ form a distinct feature at the top of the valence band which becomes partially depopulated for x>0. A (inter-atomic) and B (intra-atomic) transitions are indicated with arrows at the bottom. (d) $La_{8-n}Sr_nFe_8O_{24}$ one-electron DOS calculated using B3LYP and embedded cluster. (e) Low-energy region of the optical absorption spectra calculated using TDDFT and B3LYP; green, red, and grey colors indicate contributions due to A and B type transitions, along with their sum, respectively.

Table 1. LSFO film thickness (estimated from growth rate), calculated x value (see Supporting Information), c-axis lattice parameter [Fig. 1(c)], onsets of optical transitions A and B [Fig. 4(a)], and the valence band maxima (Fig. S3)

| Composition | Thickness (nm) | Calculated x value | c-axis lattice parameter (Å) | Onset of Transition A (eV) | Onset of Transition B (eV) | VBM (eV) |
|---|---|---|---|---|---|---|
| x = 0 | 10 | .. | 3.974 | 2.30 | .. | 1.02 |
| x = 0.2 | 30 | 0.21 | 3.955 | 1.97 | 0.49 | 0.84 |
| x = 0.37 | 30 | 0.35 | 3.943 | 1.53 | 0.34 | 0.70 |
| x = 0.5 | 30 | 0.53 | 3.915 | 1.40 | 0.23 | 0.49 |
| x = 0.8 | 30 | 0.77 | 3.871 | 1.26 | 0.10 | 0.17 |
| x = 1 | 30 | .. | 3.831 | 1.05 | 0 | 0 |

**Supporting information**

**Computational modeling**

To generate structures for computing LSFO optical absorption spectra, we first performed periodic model DFT (PBEsol) simulations of the 40-atom (2×2×2) perovskite cell $La_{8-n}Sr_nFe_8O_{24}$, where 8–$n$ La and $n$ Sr atoms, respectively, were randomly arranged in the A sublattice. For example, structures generated for $n$ = 0, 1, 7, 8, together with the corresponding distribution of spin-densities, are shown in Fig. S1(a). The results of these calculations suggest that three types of changes occur in LSFO with increasing concentration of Sr. First, surfaces of the constant spin-density change from the sphere-like shape to the cube-like shape. (Here, for simplicity, we adopted the G-type spin arrangements for all cases.) This change is consistent with a gradual change of $Fe^{3+}$ oxidation states to $Fe^{4+}$. Second, the character of the local octahedral distortions changes as the system transforms from $LaFeO_3$ to $SrFeO_3$. This change is apparent in Fig. S1(a). In addition, the dependence of the optimal supercell volume on the Sr concentration, shown in Fig. S1(b), points to a discontinuity associated with this structural phase transition. Finally, the one-electron densities of states (DOS) shown in Fig. S1(c) evolve such that the intensity of the band located between –1 and 0 eV decreases. As discussed in the main text, this band is associated with Fe $3d$ $e_g$ – O $2p$ hybridized orbitals; it's depletion is consistent with the gradual conversion of $Fe^{3+}$ to $Fe^{4+}$ with increasing Sr concentration. Simultaneously, the band gap between the occupied (black) and unoccupied (red) states decreases, consistent with the experimentally observed shift of the optical absorption onset to lower energies.

To construct the optical absorption spectra, we first represent a fragment of LSFO in the form of a finite cluster that is modeled quantum mechanically (QM cluster for short) embedded into the potential produced by the rest of the infinite lattice. Then, optical excitation energies are calculated using TDDFT, as implemented in the NWChem code, represented in the form of a Gaussian-type function with the full width at half maximum of 0.1 eV and superimposed.

The procedure for constructing the embedded cluster and the corresponding embedding potential is illustrated in Fig. S1(d). The embedding potential includes two contributions: (i) the long-range electrostatic potential due to the lattice ions, and, (ii) the short-range confinement potential to avoid artificial polarization of electron density at the boundary of the quantum cluster. To reproduce the electrostatic potential, we complement the periodic supercell with auxiliary point charges so as to eliminate dipole, quadrupole, and hexadecapole moments of the supercell using an approach described elsewhere [1,2]. Here all ions of the supercell are represented using point ions with the corresponding formal charges (in atomic units: +2 for Sr, +3 for La, –2 for O), except Fe, for which the average ionic charge consistent with $La_{8-n}Sr_nFe_8O_{24}$ composition was used. Then, a spherical fragment of LFSO, shown in the middle panel of Figure 1(d), was constructed from such modified supercells. The auxiliary charges cancel each other out exactly in the inner part of the sphere. The charges remaining on the outer part of the sphere reproduce the lattice contribution of the Madelung potential.

For consistency, QM clusters for each $n$ include 8 Fe atoms of the corresponding supercell, 7 A site atoms, either La or Sr depending of $n$, and 36 O atoms nearest to the cluster center. For all cations we discarded the most diffuse basis function to avoid artificial electron delocalization in the excited states. The next shell of A and B sublattice atoms is also included in the QM cluster, as shown in the right panel of Fig. 1(d). The purpose of these ions is to provide the confinement potential for the electronic states in the inner part of the cluster. These atoms are represented using large core effective core pseudopotentials (ECP) [3,4]. Due to the similarities of the ionic radii, we used the short-range parts of the Al and Na ECP to represent the confinement potential due to Fe and La, respectively; the Coulomb component of these potentials was adjusted to produce the correct electrostatic contribution.

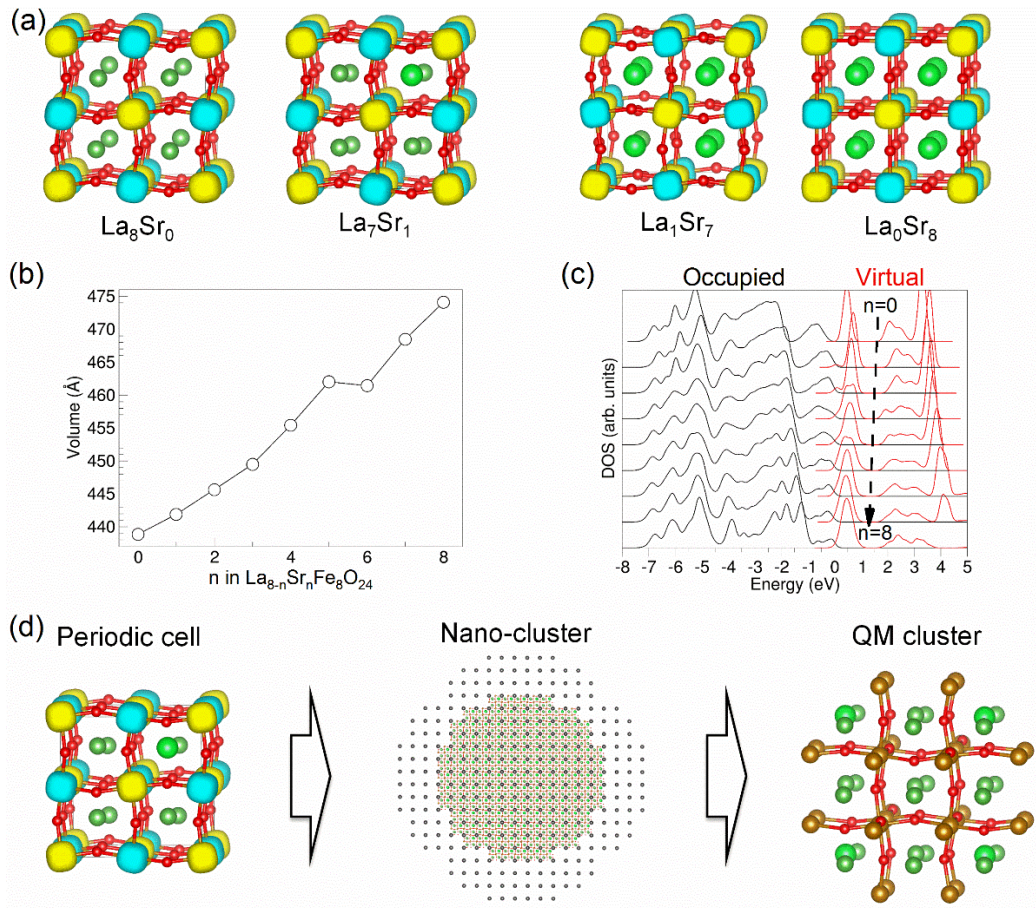

**Figure S1**. Periodic model approach (a-c). (a) 40-atom (2×2×2) perovskite cell $La_{8-n}Sr_nFe_8O_{24}$, where 8–$n$ La and n Sr atoms, respectively, were randomly arranged in the A sublattice. The La (small green), Sr (large green) and O (red) atoms are shown explicitly. Fe atoms (brown) are hidden behind the surfaces of constant spin-density; spin-up and spin-down electronic configurations are indicated with yellow and turquoise colors. (b) Supercell volume as a function of $n$ in $La_{8-n}Sr_nFe_8O_{24}$, calculated using PBEsol functional. (c) Density of states (DOS) evolution with $n$. Occupied and virtual DOS are shown with black and red lines, respectively. (d) Embedded cluster approach: optimized LSFO periodic cells (left) were used to construct finite spherical region representations of the LSFO lattice (center) which provides the embedding potential for the quantum mechanically treated cluster (right) located at the center of the sphere.

**Experimental details and additional data**

To determine x in $La_{1-x}Sr_xFeO_3$, we used a simple procedure requiring no sensitivity factors or photoemission cross sections, in which a pure LFO film was our standard for a ferrite perovskite lattice with 100% of the A sites occupied by La [5]. The procedure involves first measuring the La $3d_{5/2}$-to-O $1s$ peak area ratio for the LSFO film set. Normalization using the O $1s$ peak area removes any spurious intensity differences in the spectra measured for different samples on different dates caused by the gradual loss of x-ray intensity over time due to Al anode degradation. The Sr mole fraction ($x$) can be related to the peak areas ($A$) by,

$$x = 1 - \frac{(A_{La3d5/2}/A_{O1s})_{LSFO}}{(A_{La3d5/2}/A_{O1s})_{LFO}} \tag{S1}$$

The deduced x values for the LSFO film set are shown in Table 1, and they verify that the actual film stoichiometry is within ~10% of the target stoichiometry based on the La and Sr fluxes.

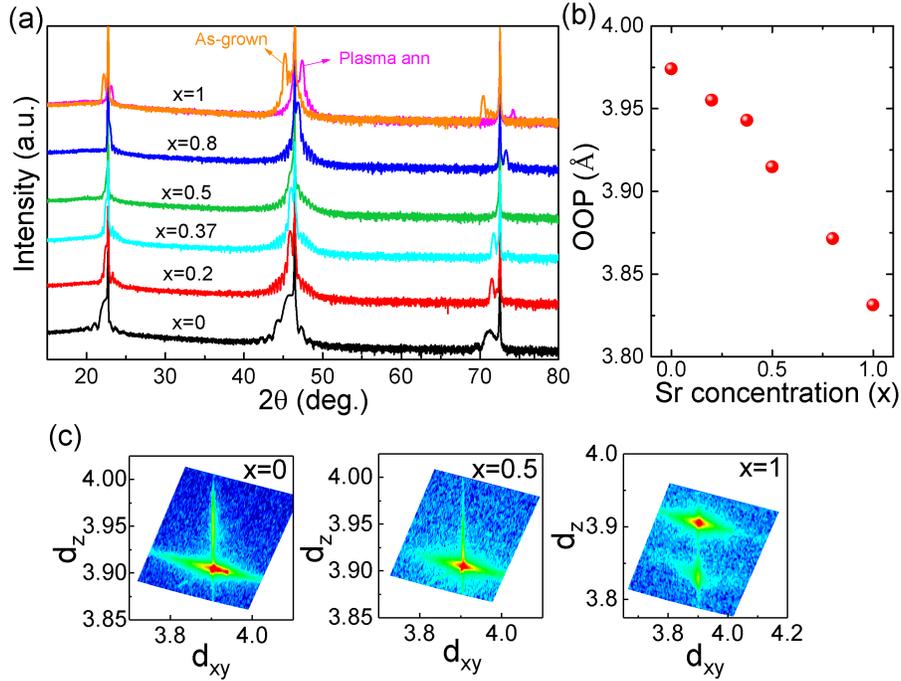

**Figure S2**. (a) Extended XRD θ-2θ patterns for the LSFO/STO film set, from which the presence of single-phase LSFO films with a c-axis orientation is deduced. All LSFO films except the as-grown SFO film are treated by annealing in activated oxygen, as described in the paper. (b) Out-of-plane lattice parameters (OOP) for LSFO films as a function of x. (c) RSMs taken near the (103) reflection for representative (x = 0, 0.5 and 1) LSFO films on STO(001) substrates.

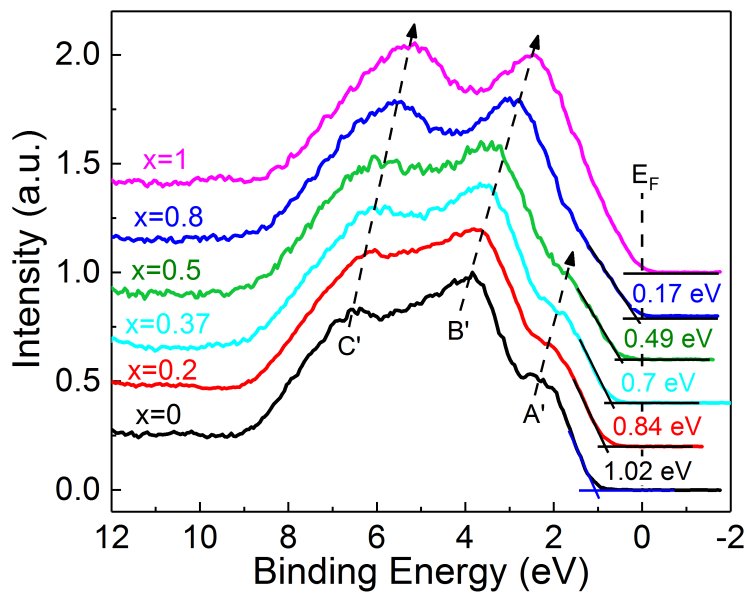

**Figure S3**. XPS VB spectra for La$_{1-x}$Sr$_x$FeO$_3$. The valence band maximum (VBM) shifts toward E$_F$ with increasing x. For LFO, three features denoted as A′, B′, and C′ can be observed. A gap (absence of finite DOS at E$_F$) was seen for all LSFO films except SFO (x = 1), indicating that SFO is metallic, consistent with the transport measurements shown in Fig. S4. Structures A′- C′ move toward E$_F$ upon increasing x. These shifts are in good agreement with the core-level shifts shown in Fig. 3(a). Structure A′ becomes weaker with increasing x, indicating that the holes are doped into the occupied e$_g$ state of Fe 3$d$ on the top of VB.

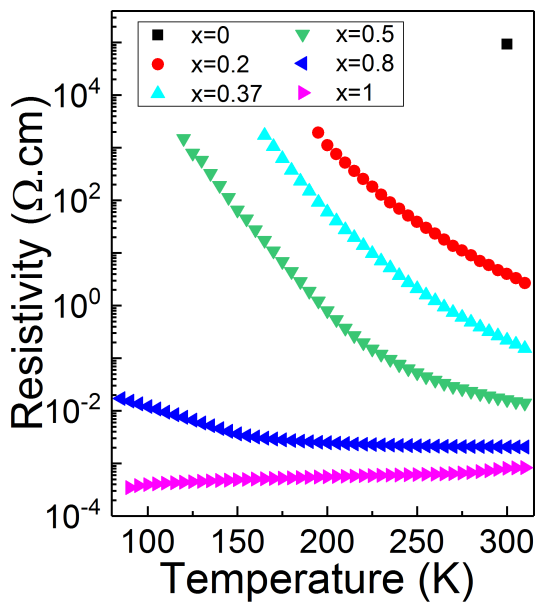

**Figure S4**. Temperature dependent electrical resistivity of the LSFO films after annealing in oxygen plasma. All LSFO films were made using the same conditions as the samples for the ellipsometry and photoemission measurements. Except the metallic behavior displayed in SFO film, all other LSFO films show semiconducting or insulating behavior, consistent with the optical measurements shown in Fig. 4(a).